\documentclass{sigchi}

\toappear{\small Permission to make digital or hard copies of all or part of this work for personal or classroom use is granted without fee provided that copies are not made or distributed for profit or commercial advantage and that copies bear this notice and the full citation on the first page. To copy otherwise, or republish, to post on servers or to redistribute to lists, requires prior specific permission and/or a fee. \newline   \emph{WebSci'13}, May 1 -- May 5, 2013, Paris, France.\\ ACM 978-1-4503-1889-1.}
\usepackage{balance}  % to better equalize the last page
\usepackage{graphics} % for EPS, load graphicx instead
\usepackage{times}    % comment if you want LaTeX's default font
\usepackage{url}      % llt: nicely formatted URLs

\usepackage{subcaption}

\usepackage{booktabs}
\usepackage{ctable}
\usepackage{multirow} % multirow support for tables
\usepackage{rotating}

\usepackage{array}

\newcolumntype{P}[1]{>{\centering\raggedright\arraybackslash}p{#1}}

\renewcommand{\arraystretch}{.924}
\newcommand{\otoprule}{\midrule[\heavyrulewidth]}

\makeatletter
\def\url@leostyle{%
  \@ifundefined{selectfont}{\def\UrlFont{\sf}}{\def\UrlFont{\small\bf\ttfamily}}}
\makeatother
\def\pprw{8.5in}
\def\pprh{11in}

\usepackage[pdftex]{hyperref}
\hypersetup{
pdftitle={WebScience Conference Proceedings Format},
pdfauthor={LaTeX},
pdfkeywords={WebScience, proceedings, archival format},
bookmarksnumbered,
pdfstartview={FitH},
colorlinks,
citecolor=black,
filecolor=black,
linkcolor=black,
urlcolor=black,
breaklinks=true,
}

\usepackage{hyperref}

\begin{document}

\title{Semantic Tagging on Historical Maps}

% Note that submissions are blind, so author information should be omitted
\numberofauthors{4}
\author{
  \alignauthor Bernhard Haslhofer\\
    \affaddr{Cornell University, Information Science}\\
    \email{bh392@cornell.edu}
  \alignauthor Werner Robitza\\
    \affaddr{University of Vienna, Computer Science}\\
    \email{werner.robitza@univie.ac.at}
  \alignauthor Carl Lagoze\\
    \affaddr{University of Michigan, School of Information}\\
    \email{clagoze@umich.edu}
  \alignauthor Francois Guimbretiere\\
    \affaddr{Cornell University, Information Science}\\
    \email{francois@cs.cornell.edu}
}

% Teaser figure can go here
%\teaser{
%  \centering
%  \includegraphics{Figure1}
%  \caption{Teaser Image}
%  \label{fig:teaser}
%}

\maketitle

\begin{abstract}

% Problem
Tags assigned by users to shared content can be ambiguous. As a possible solution, we propose \emph{semantic tagging} as a collaborative process in which a user selects and associates Web resources drawn from a knowledge context. We applied this general technique in the specific context of online historical maps and allowed users to annotate and tag them.
% Applied method
To study the effects of semantic tagging on tag production, the types and categories of obtained tags, and user task load, we conducted an in-lab within-subject experiment with 24 participants who annotated and tagged two distinct maps.
% Important Results
We found that the semantic tagging implementation does not affect these parameters, while providing tagging relationships to well-defined concept definitions. Compared to label-based tagging, our technique also gathers positive and negative tagging relationships.
% Implications
We believe that our findings carry implications for designers who want to adopt semantic tagging in other contexts and systems on the Web.

\end{abstract}


%A category including the fourth, optional field follows...


\section{Introduction}\label{sec:introduction}

% Background
Tagging is a collaborative process in which a user adds textual labels (tags) to shared content. It does not rely on static, pre-defined taxonomic structures but on dynamic, community-driven linguistic terms and conceptions~\cite{golder2006usage}. Tagging became popular with the launch of sites like Delicious and Flickr around 2005 and is now a standard feature that can be found in many social media sites. One can, for example, attach tags to media objects (Flickr, YouTube), blog entries (Blogger, WordPress, LiveJournal), news stories (Digg), posts in eLearning environments (Piazza), tweets (hashtags in Twitter), or any Web page (Delicious bookmarks, Facebook Like button). It is also gaining increasing attention in the digital humanities area, where institutions publish their collections on social media sites such as Flickr Commons\footnote{\url{http://www.flickr.com/commons}} to increase discovery and reuse of their collections and to gather descriptions and contextual information for items in their collections.

% Problems
Despite their wide-spread adoption, tagging systems still face a number of problems: a tag can be ambiguous and have many related meanings (\emph{polysemy}), multiple tags can have the same meaning (\emph{synonymy}), or the semantics of a tag might range from very specific to very general because people describe resources along a continuum of specificity~\cite{golder2006usage}. These issues are rooted in label-based nature of tags and important for system providers who want to exploit the semantics and contextual information associated with tags for resource discovery. If, for instance, a user tags a resource with ``Paris'' it is not entirely clear whether this tag means ``Paris'', the capital of France or ``Paris'', the city in the United States. Contextual information, such as the translations of the term ``Paris'' in other world-languages or its geographical location can only be determined after reconciling label-based tags with data entries in other data sources.

% Mini-literature review on semantic tagging
Mapping label-based tags to concepts defined in knowledge contexts, such as Wikipedia is a possible solution. Sigurbj\"{o}rnsson and van Zwol~\cite{Sigurbjornsson:2008vn} use string matching to map Flickr tags to WordNet semantic categories and found that 51.8\% of the tags in Flickr can be mapped. Overell et al.~\cite{Overell:2009:CTU:1498759.1498810} use concept definitions from Wikipedia and Open Directory to classify tags automatically and show that nearly 70\% of Flickr tags can be classified correctly. Also, Rattenbury et al.~\cite{Rattenbury:2007:TAE:1277741.1277762} extract place and event semantics from Flickr tags with high precision but low recall. However, in all these approaches tag semantics is determined heuristically and a-posteriori, without taking into account the user who created and assigned the tag and knows about its precise semantics.

% Our approach
To solve this problem, we propose that users associate URI-identified Web resources from a knowledge context, such as Wikipedia, as part of their tagging activity. A tagging system could suggest the label ``Paris'' as a possible tag in the user-interface, but create a link to a Web resource (e.g., \url{http://en.wikipedia.org/wiki/Paris}) in the back-end. We call this technique \emph{semantic tagging}. Different from label-based tagging, the semantics of a tag is determined by its creator at creation time. Each tag also leads to further contextual information that can be exploited for resource discovery purposes. Explicit user feedback on suggested tags results in a graph of positive and negative tagging relationships that can be used to improve tag recommendation strategies.

% Our contribution
To demonstrate the user acceptance of this approach, we implemented semantic tagging in a prototypical Web application, called Maphub. It allows users to annotate and tag zoomable high-resolution maps from the Library of Congress historical Map collection with textual comments and suggests possibly relevant tags. Our application illustrates how to design semantic tagging systems so that users can easily select from suggested semantic tags (\emph{tag recommendation}), accept or reject proposed tags (\emph{user feedback}), without ever having to interact with URIs directly (\emph{user transparency}).

Using this application, we ran an empirical evaluation to compare semantic tagging with other tagging techniques. Our central findings can be summarized as follows:

\begin{itemize}
  \item Our semantic tagging implementation does not affect tag production, the types and categories of obtained tags, and user task load, while providing tagging relationships to well-defined concept definitions.
  \item When compared to label-based tagging, our technique also gathers positive and negative tagging relationships, which can be useful for improving tag recommendation and resource retrieval.
\end{itemize}

Even though we applied the semantic tagging technique in the context of historical maps, we believe that our findings carry implications for designers who want to adopt these techniques in other contexts and systems on the Web.

\section{Semantic Tagging}\label{sec:semantic_tagging}

We now discuss the conceptual and design-related aspects of the semantic tagging technique and compare it with existing, label-based tagging design characteristics.

\subsection{Conceptual Model}

\begin{figure}
        \centering
        \begin{subfigure}[b]{0.3\textwidth}
                \centering
                \includegraphics[width=0.9\columnwidth]{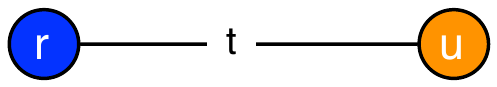}
                \caption{Label-based tagging (based on~\cite{Marlow2006tagging})}
                \label{fig:tagging_model}
        \end{subfigure}
        \begin{subfigure}[b]{0.35\textwidth}
                \centering
                \includegraphics[width=0.9\columnwidth]{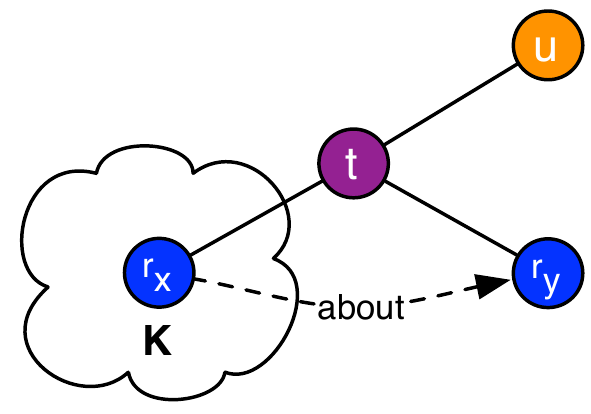}
                \caption{Semantic Tagging.}
                \label{fig:semantic_tagging_model}
        \end{subfigure}
        \caption{Label-based versus semantic tagging.}
        \label{fig:animals}
\end{figure}

In the conceptual model for label-based tagging systems introduced by Marlow et al.~\cite{Marlow2006tagging}, which is shown in Figure~\ref{fig:tagging_model}, a user $u$ assigns a tag $t$ to a resource $r$. Tags are represented as labeled edges that connect users and resources but do not carry or refer to any additional contextual information. Both resources and users may be connected to other nodes, since there may be links between Web pages and users may belong to social networks. Label based tagging systems can allow for multiplicity of tags around resources (\emph{bag-model}) or deny tag repetitions (\emph{set-model})~\cite{Marlow2006tagging}.

Semantic tagging, which is shown in Figure~\ref{fig:semantic_tagging_model}, extends this model by representing a tag $t$ as a qualified relationship between two resources: $r_x$ is the resource identifying and defining the semantics of a tag (e.g., \url{http://en.wikipedia.org/wiki/Paris}), and $r_y$ is the resource being tagged (e.g., a photo taken in Paris). The former is defined within a knowledge context $K$ and can carry textual labels (e.g., ``Paris'') and additional context information (e.g., Paris is a city in France). Possible knowledge contexts are online encyclopedias such as Wikipedia, place name registries such as GeoNames, structured Web data sources such as Freebase\footnote{\url{http://www.freebase.com/}}, domain-specific Web vocabularies or gazetteers, or any other Linked Data source providing suitable concept definitions. An explicit, qualified semantic tagging relationship also implies an \emph{about} relationship between the involved resources, meaning that $r_x$ is about $r_y$ if they are connected by a user via a semantic tagging relationship.

Since semantic tags can also be represented as first-class URI-identified Web resources, the resulting model is not label- or set-based but \emph{graph-based}, with different types of nodes (users, resources) being connected to each other. This enables multiplicity and aggregation of tags not only around resources but also around users and user groups, which can be exploited for graph-based tag recommendation and user-based collaborative tag filtering~\cite{Jaschke:2007ys}.

We believe that an information system implementing semantic tagging should allow users to easily select from suggested tags, accept or reject proposed tags, without ever having to interact with URIs. Therefore we will now continue discussing the following design aspects in more detail: \emph{tag recommendation}, \emph{user feedback}, and \emph{user transparency}.

\subsection{Tag recommendation}

Marlow et al.~\cite{Marlow2006tagging} distinguish between three main categories existing systems fall into: \emph{blind} tagging, where a user cannot view tags assigned to the same resource by other users, \emph{viewable} tagging, where users sees tags associated with a resource, and \emph{suggestive} tagging, where the system suggests possible tags to the user. Suggestive tagging systems can derive tags from existing tags by the same or other users or gather them from a resource's context.

% tag recommendation strategies
Following this classification, we propose semantic tagging as a special form of suggestive tagging, where tag resources are recommended from a given knowledge context, based on the context of any resource that is part of the semantic tagging graph. As in other suggestive tagging systems (see~\cite{gupta2010survey}), tag recommendation strategies can consider the content (e.g., image file) or context (e.g., metadata, other tag resources) of a resource. If the applied knowledge context follows a graph structure, it is also possible to apply graph-based recommendation strategies for tag resource proposals. When, for instance, a system proposes the semantic tag ``Paris'', it could also propose related resources such as ``France'', and ``Eiffel Tower'' if these concepts are semantically connected in the underlying knowledge context---as it is the case in Wikipedia. In Maphub, for example, we recommend semantic tags based on the text users are entering while they are authoring annotations on historical maps.

% tag recommendation solutions
Semantic tag suggestion can be implemented by calling named entity recognition services that link things mentioned in plain text to Web resources, such as Wikipedia Miner\footnote{\url{http://wikipedia-miner.cms.waikato.ac.nz/}} or DBPedia Spotlight\footnote{\url{https://github.com/dbpedia-spotlight/dbpedia-spotlight/wiki}}.

\subsection{User feedback}

% positive relationships
Adding a label-based or semantic tag to a given resource usually means that the tag is somehow about or describes the resource, at least within the context of the tag creator. If a user applies the tag ``Paris'' to an image it is assumed that Paris is somehow about that image. Thus, an intrinsic assumption of existing tagging models is that relationships between tags and resources have positive connotations.

\begin{figure}
  \centering
  \includegraphics[width=0.7\columnwidth]{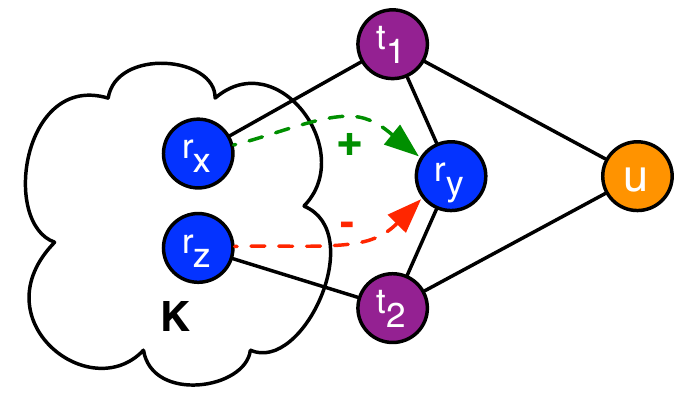}
  \caption{Semantic tags forming a graph with positive and negative relationships.}
  \label{fig:pos_neg_rels}
\end{figure}

% negative relationships
However, with tags becoming first-class resources describing a qualified relationship between resources, one can also capture negative relationships: when the system recommends a set of possibly relevant (semantic) tags and the user accepts one of them, it can infer a positive tagging relationships. However, the system could also capture the non-accepted or explicitly rejected tags and interpret them as negative tagging relationships, as illustrated in Figure~\ref{fig:pos_neg_rels}. An explicitly rejected tag ``Berlin'' on an image showing Paris is an example for such a negative relationship.

% exploiting positive and negative relationships
Qualified semantic tagging relationships carrying positive and negative weights can easily be transformed into a bipartite graph of positive (accept) and negative (reject) ``about'' relationships between semantic tags and tagged resources. From this graph, one can directly derive relevance judgments for given pairs of Web resources and build gold standards, which are required for subsequent information retrieval tasks.
%In a related study~\cite{Murnane:2013uq}, for instance, we show that modeling user interest with respect to a knowledge context (e.g., Wikipedia) can bring up to 20\% gains in performance beyond state-of-the-art Named Entity detection techniques.
Positive and negative relevance judgments can also be exploited by active learning techniques~\cite{Settles:2010vn} to improve tag recommendation models.

\subsection{User transparency}

The World Wide Web uses HTTP URIs to unambiguously identify Web resources, such as the Wikipedia article about Paris. However, URIs are opaque strings that do not necessarily carry any semantics. While the design choice in Wikipedia was to use-human readable URIs (e.g., \url{http://en.wikipedia.org/wiki/Paris}, other sources do not follow this approach. In the GeoNames knowledge context, for instance, Paris is identified by a URI with a numeric path element \url{http://www.geonames.org/2988507}. Such a URI syntax is hard to remember for human end-users and might lead to errors when being transcribed manually.

Therefore, semantic tagging systems should hide the technical aspects of this approach
%underneath the user-interface
and follow the design of existing suggestive tagging interfaces: they should neither display nor prompt users to input HTTP URIs, but suggest labels and maintain internal, user-transparent mappings between labels and their corresponding resources. For example, instead of displaying a semantic tag URI for Paris the system should present labels such as ``Paris''.

This of course requires that the knowledge context also provide human-readable labels for resource definitions, which is common practice in real-world data sources. In the case of Wikipedia one can, for instance, extract the article's title (``Paris'') directly from the Web page or rely on DBpedia~\cite{Auer:2007zr}, which provides structured data extracted from Wikipedia.

\section{Maphub}\label{sec:maphub}

\begin{figure}
  \centering
  \includegraphics[width=1\linewidth]{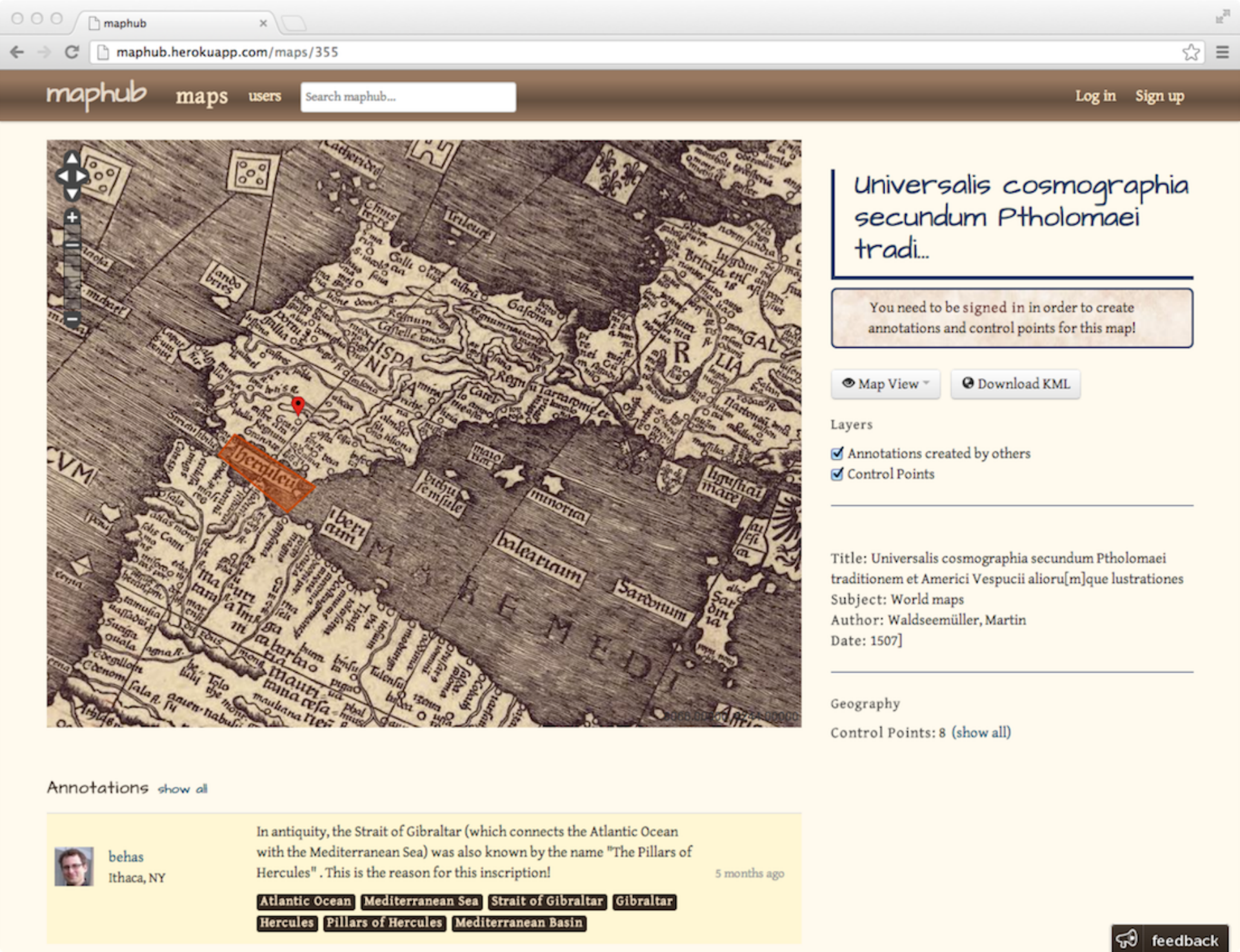}
  \caption{Maphub prototype screenshot.}
  \label{fig:maphub}
\end{figure}

% Maphub in general
Maphub\footnote{\url{http://maphub.github.com}} is an open source web application that allows users to create annotations on historical maps and implements semantic tagging as a feature. Currently, the system supports two main use cases: (i) georeferencing maps by creating \emph{control points} and (ii) commenting on maps or map regions by creating \emph{annotations}. By adding control points, users link points on a historic map to named real-world locations whose coordinates correspond to that point. By creating annotations they can add their comments and express their knowledge about the map or a specific map region. Figure~\ref{fig:maphub} shows a screenshot of the system.

A first prototype\footnote{\url{http://maphub.herokuapp.com}} has been bootstrapped with a set of around 6,000 digitized high-resolution historical maps from the Library of Congress' Map Division. It allows users to retrieve maps either by browsing or searching over available metadata and user-contributed annotations and tags.

% How does Maphub implement Semantic Tagging
Semantic tagging is part of Maphub's annotation feature: to create an annotation, users mark up regions on the map with geometric shapes such as polygons or rectangles. Once the area to be annotated is defined, they are asked to tell their stories and contribute their knowledge in the form of textual comments. While users are composing their comments, Maphub periodically suggests tags based on either the text contents or the geographic location of the annotated map region. Suggested tags appear below the annotation text. The user may \emph{accept} tags and deem them as relevant to their annotation or \emph{reject} non-relevant tags. Unselected tags remain \emph{neutral}.

\begin{figure}
  \centering
  \includegraphics[width=1\linewidth]{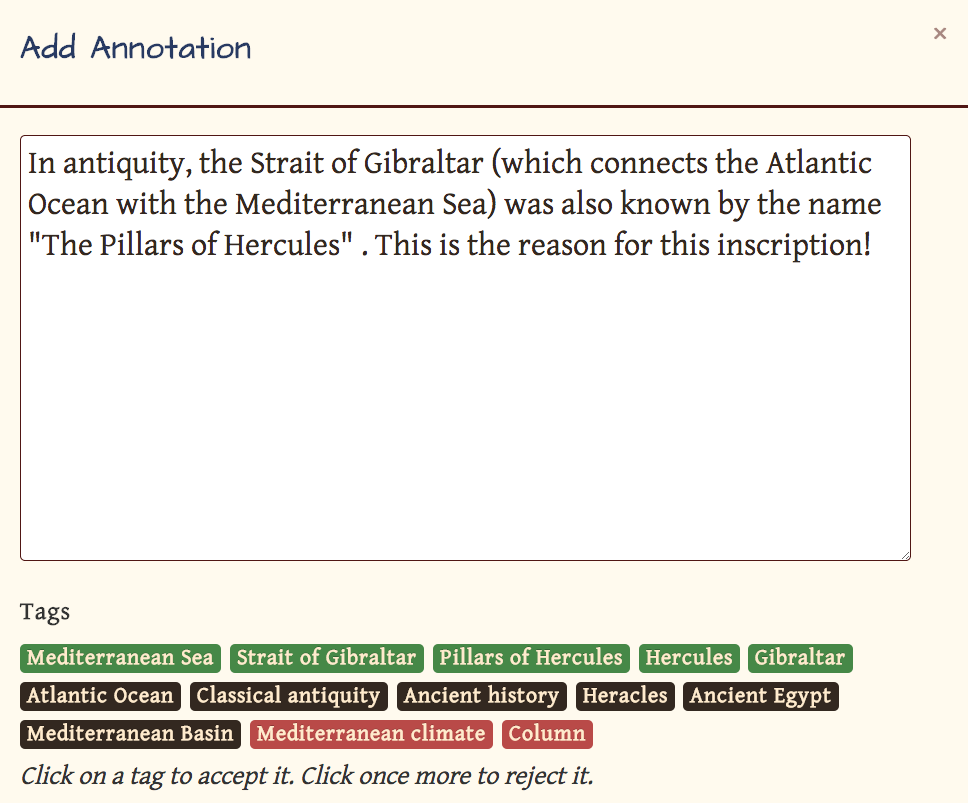}
  \caption{Detailed annotation screenshot.}
  \label{fig:annotation}
\end{figure}

The screenshot in Figure~\ref{fig:annotation} shows an example user annotation created for a region covering the Strait of Gibraltar. While the user entered a free-text comment related to the naming of the area, Maphub queried an instance of Wikipedia Miner to perform named entity recognition on the entered text and received a ranked list of Wikipedia resource URIs (e.g., \url{http://en.wikipedia.org/wiki/Mediterranean_sea}) in return. URIs should not be exposed to the user, so Maphub displays the corresponding Wikipedia page titles instead (e.g., ``Mediterranean Sea''). Since page titles alone might not carry enough information for the user to disambiguate concepts, Maphub offers additional context information: the short abstract of the corresponding Wikipedia article is shown when the user hovers over a tag.

Since users also enter control points, we can compute an approximate mapping from the raster map image to real-world projections (e.g., Spherical Mercator) and suggest semantic tags also based on the geographic region of the annotation using Web services such as GeoNames\footnote{\url{http://www.geonames.org/}}. Once the user has entered text, the rectangular geographical boundary box for the marked area is calculated and transformed from map coordinates to real-world coordinates. We pass these boundaries to the GeoNames API, which in turn delivers a list of geo-tagged Wikipedia resources contained within that boundary. Similar to semantic tags obtained through named entity recognition, Wikipedia page titles and excerpts are shown to the user as additional context information.

Once tags are displayed, users may mark them as relevant for their annotation by clicking on them once, which turns the labels green. Clicking once more rejects the tags, and clicking again sets them back to their (initial) neutral state. In Figure~\ref{fig:annotation}, the user accepts five tags and actively prunes two tags that are not relevant in the context of this annotation.

% Annotation sharing
Sharing collected annotation data in an interoperable way was another major development goal. Maphub is an early adopter of the Open Annotation model, which is currently specified in the W3C Open Annotation working group\footnote{\url{http://www.w3.org/community/openannotation/}}. It demonstrates how to apply that model in the context of digitized historic maps and how to expose comments as well as semantic tags. As described in the Maphub API documentation\footnote{\url{http://maphub.github.com/api}}, each annotation becomes a first class Web resource that is dereferencable by its URI and therefore easily accessible by any Web client. In that way, while users are annotating maps, Maphub not only consumes data from global data networks---it also contributes data back.

\section{Empirical Evaluation}\label{sec:methodology}

In an in-lab experiment we studied how semantic tagging, compared to other tagging techniques, affects tag usage, the types and categories of added tags, and user satisfaction.

\subsection{Methodology}

Our experiment follows a within-subject design, with four tag creation methods as the main varying condition:

\begin{enumerate}
  \item \textbf{Label-based tagging (LT):} the user enters comma-separated label-based tags similar to social media tagging systems such as Flickr or YouTube.
  \item \textbf{Suggestive tagging (ST):} The system randomly selects and suggests label-based and semantic tags that were entered by the same or other users before.
  \item \textbf{Semantic tagging (SMT):} Tags are generated and suggested based on the text and geographical area of an annotation.
  \item \textbf{Semantic tagging with Context Display (SMT-CTX):} Tags are generated and suggested as in the SMT condition, but the user is shown additional contextual information (Wikipedia page abstract) when hovering over it.
\end{enumerate}

In order to avoid learning effects, we counterbalanced the four conditions following a balanced Latin Square design. This guarantees that each condition appears equally often in the order of condition assignments to participants and assures that each condition precedes and follows each of the other conditions equally often. In all four conditions, already existing tags were not shown in the tagging interface. The user could, however, browse existing annotations and view already assigned tags.

While the tagging condition with levels LT, ST, SMT, and SMT-CTX was our independent variable, the main dependent variables were the number of accepted and rejected tags added per annotation. We also measured the effect on the users' task load using a simplified NASA Task Load Index (TLX) questionnaire, which lets participants rate the mental, physical and temporal demand of a task on a 7-point Likert scale. It also records how successful they think they were in accomplishing what they were asked to do, as well as the amount of effort needed and the level of frustration. Participants also ranked the conditions they had seen according to intuitiveness, influence on annotation text, mental effort, and overall usefulness.

In order to test each participant with four different annotation interfaces, each implementing a different tagging condition, we implemented our manipulations in an experimental branch of the Maphub application and deployed it separately\footnote{\url{http://maphub-experiment.herokuapp.com}} from the production system.

\subsection{Procedure}

Two student experiment administrators alternately conducted the experiment sessions with one participant at a time. After signing an experiment consent form and filling out the pre-test survey, an experiment administrator opened the experimental Maphub deployment in a Web browser and demonstrated Maphub's features and functionality by creating the sample annotation shown in Figure~\ref{fig:annotation}. The participant was then asked to create a user profile in Maphub and to repeat the following steps four times:

\begin{enumerate}
  \item Choose one of the two maps and identify a region you are familiar with.
  \item Create an annotation for that region and add tags (note that the tagging user interface will change).
  \item Fill out the NASA-TLX questionnaire in the post-test survey.
\end{enumerate}

Participants were asked to speak out loudly while working with the Maphub system. The experiment administrators took notes. At the end of each session, the participants ranked the conditions they had seen, entered additional feedback into the post-test survey form, and signed a payment confirmation sheet.

\subsection{Participants and Maps}

26 participants were recruited via Cornell-internal mailing lists. They were asked to fill out a pre-test survey, create annotations and tags on a selection of two maps, and report their experiences in a post-test survey. A compensation of \$15 was paid for their approximately 45 minutes effort of taking part in the experiment. The data from two participants was discarded because of technical difficulties.
% We also removed annotations that were created for demo purposes by our two experiment administrators.

This left us with data from 24 participants who were mainly students from the field of information science: 16 male and 8 female with a mean age of 23. When being asked about their reading habits, the majority of participants responded that they actively read about social media, economics and politics at least once a week. Half of them never read about geographical information services or cartography (54\% each) and their interest in topics such as geography and history was low. We can thus assume that our participants have only marginal previous experience in these fields and can therefore be considered as non-experts.

In the pre-test survey, we also asked the participants about their familiarity with the two main Maphub features used in this study: tagging and annotating. As expected, tagging is well known: the majority (77\%) responded they used tagging at least once a week. The term ``tagging'' was clear to everyone. The concept of annotations, however, was unknown to two participants, with more than half of the users (60\%) never having used annotations at all.

We tried to minimize the confounding variables of this experiment by allowing annotations only on two maps: (i) the Waldseem\"{u}ller map, which is currently housed at the Library of Congress and known as the first map to use the word ``America'', and (ii) a map from the 18th century that shows the east coast of the United States. We chose the first map due to its popularity and the second because we assumed that participants studying at Cornell would be familiar with that region.

\subsection{Data Coding}

% Tag type and category coding
In order to determine the \emph{type} of a tag, we followed Sen et al.~\cite{sen2006tagging} and manually coded the 221 tags in our dataset into one of the following two classes: \emph{factual tags} identifying ``facts'' that most people would agree apply to a given map, and \emph{personal tags}, which are tags that have an intended audience of the tag applier themselves. We also categorized tags into the following semantic categories, as proposed by Sigurbj\"{o}rnsson et al.~\cite{Sigurbjornsson:2008vn}: \emph{locations}, \emph{artifacts or objects}, \emph{people or groups}, \emph{actions or events}, and \emph{time}. If tags were incomprehensible or did not fit in a class or category, the tag was coded as class \emph{other}. Each tag was coded by two people with a Cohen's Kappa inter-rater reliability of $\kappa = 0.85$ for tag types and $\kappa = 0.79$ for tag categories. When coders differed, they discussed the tag and reached a consensus.

\section{Results}\label{sec:results}

Our dataset contains data collected from 24 participants, where each created 4 annotations under 4 different tagging conditions (LT, ST, SMT, SMT-CTX). This gives us 96 annotations in total, each carrying zero or more accepted and/or rejected tags. The Waldseem\"{u}ller map received 45 annotations, the map from the 18th century showing the US east cost 51 annotations. In total we received 221 tags, the majority (195) being distinct. Table~\ref{tab:basic_stats} shows the most frequent tags ($freq > 1$) added under each condition. All other tags were assigned just once.

\begin{table}
  \centering
  \resizebox{0.79\columnwidth}{!}{
  \begin{tabular}{P{1.5cm}P{5cm}}
      Condition & Most Frequent Tags \\ \otoprule%
      \textbf{LT} & Ithaca (6), Cornell University (3), New York (2) \\ \midrule%
      \textbf{ST} & Culture (2), historical differences (2), New Jersey (2) \\ \midrule%
      \textbf{SMT} & India (3), Japan (3), New York City (2), United States (2) \\ \midrule%
      \textbf{SMT-CTX} & Capital City (2), Geographical pole (2), Hudson River (2), North America (2), Pennsylvania (2), United States (2) \\ 
      \bottomrule%
  \end{tabular}
  }
  \caption{Most frequent tags added under each tagging condition.}
  \label{tab:basic_stats}
\end{table}

\begin{figure}
    \centering
    \includegraphics[width=01\columnwidth]{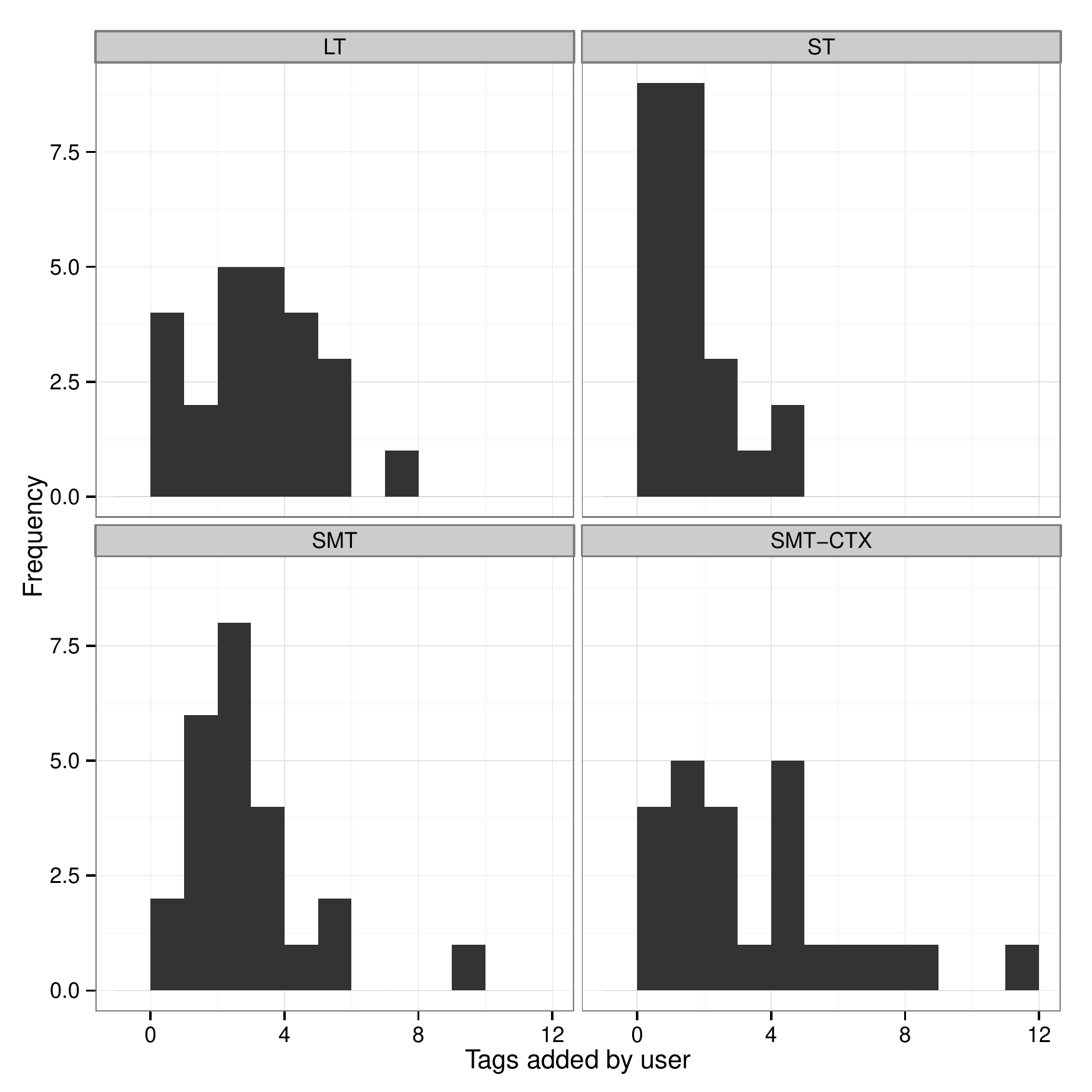}
    \caption{Tag frequencies under each tagging condition.}
    \label{fig_tag_frequencies}
\end{figure}

The number of accepted tags added as part of a single annotation, which is shown in Figure~\ref{fig_tag_frequencies}, indicates a left-skewed distribution for all four conditions. This means that either no tag was assigned or a small number of tags was applied frequently. The majority of tags was applied only a few times or just once. It also shows that the participants in our experiment created less annotations without tags under conditions SMT or SMT-CTX than under conditions LT and ST. The number of annotations without tags was highest under condition ST, which proposed tags based on already existing tags.
%Annotations created under the semantic tagging conditions (SMT and SMT-CTX) resulted in less annotations without tags.

\subsection{Effect on Tag Production}

A repeated measures analysis of variance computed on the mean number of accepted tags indicates that there is a significant difference between the four tagging conditions in their capacity to receive accepted tags as part of annotations, $F(3,69) = 6.83, p < .01$. Posthoc pairwise comparisons using paired t-tests with Bonferroni correction revealed that the mean number of accepted tags in condition ST is significantly different from LT ($p < .01$), SMT and SMT-CTX ($p < .05$). The same analysis applied on rejected tags shows that there was a significant difference between the four conditions in their capacity to receive rejected tags, $F(3,69) = 19.78, p < .01$. The mean number of rejected tags in ST is significantly different from all other conditions ($p < .01$) and LT is significantly different from SMT ($p < .05$).

\begin{figure}
    \centering
    \includegraphics[width=1\columnwidth]{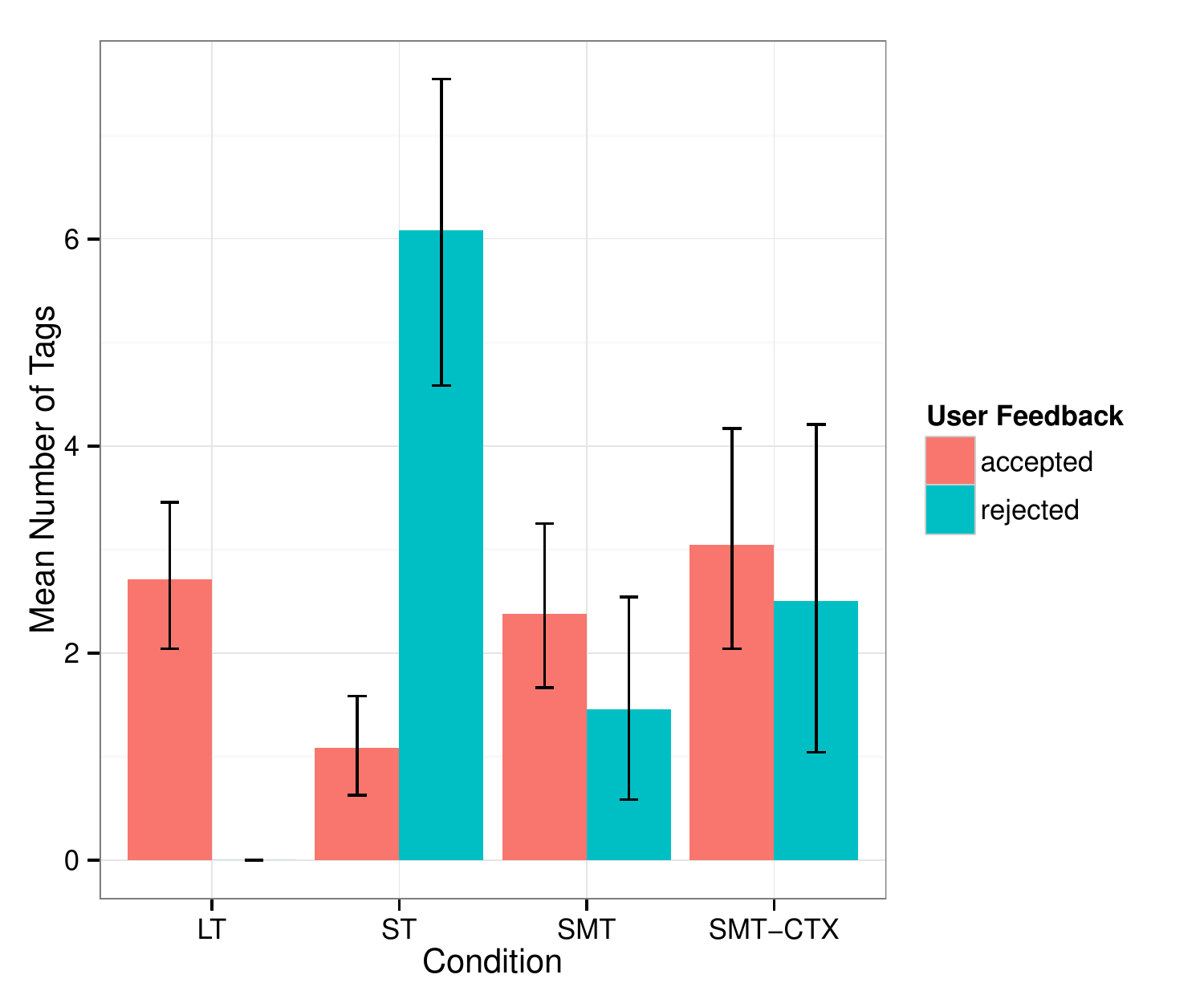}
    \caption{Effect on mean number of accepted and rejected tags per annotation.}
    \label{fig_number_of_tags}
\end{figure}

These tendencies can also be observed in Figure~\ref{fig_number_of_tags}, which shows the mean number of accepted and rejected tags per annotation with their 95\% confidence intervals. The means were 2.71 (LT), 1.08 (ST), 2.37 (SMT), and 3.04 (SMT-CTX). For rejected tags the means were 0 (LT), 6.08 (ST), 1.46 (SMT), and 2.50 (SMT-CTX). This shows that users added roughly the same number of tags under the semantic tagging conditions SMT and SMT-CTX as under the label-based tagging condition LT. It also shows that the semantic tagging techniques (SMT, SMT-CTX), which are both suggestive, lead to less rejected tags than suggestive tagging based on already added tags (ST). Showing Wikipedia article abstracts as contextual information (SMT-CTX) leads to a higher mean number of accepted tags but did not make a significant difference compared to techniques LT and SMT. However, it is important to note that the means shown in this figure depend on the number of tags proposed on the user interface. If, for instance, we showed only three tags under condition ST we would restrict the maximum possible mean to this number. The maximum number of tags we displayed in this Maphub experiment was 15.

\begin{figure}
    \centering
    \includegraphics[width=1\columnwidth]{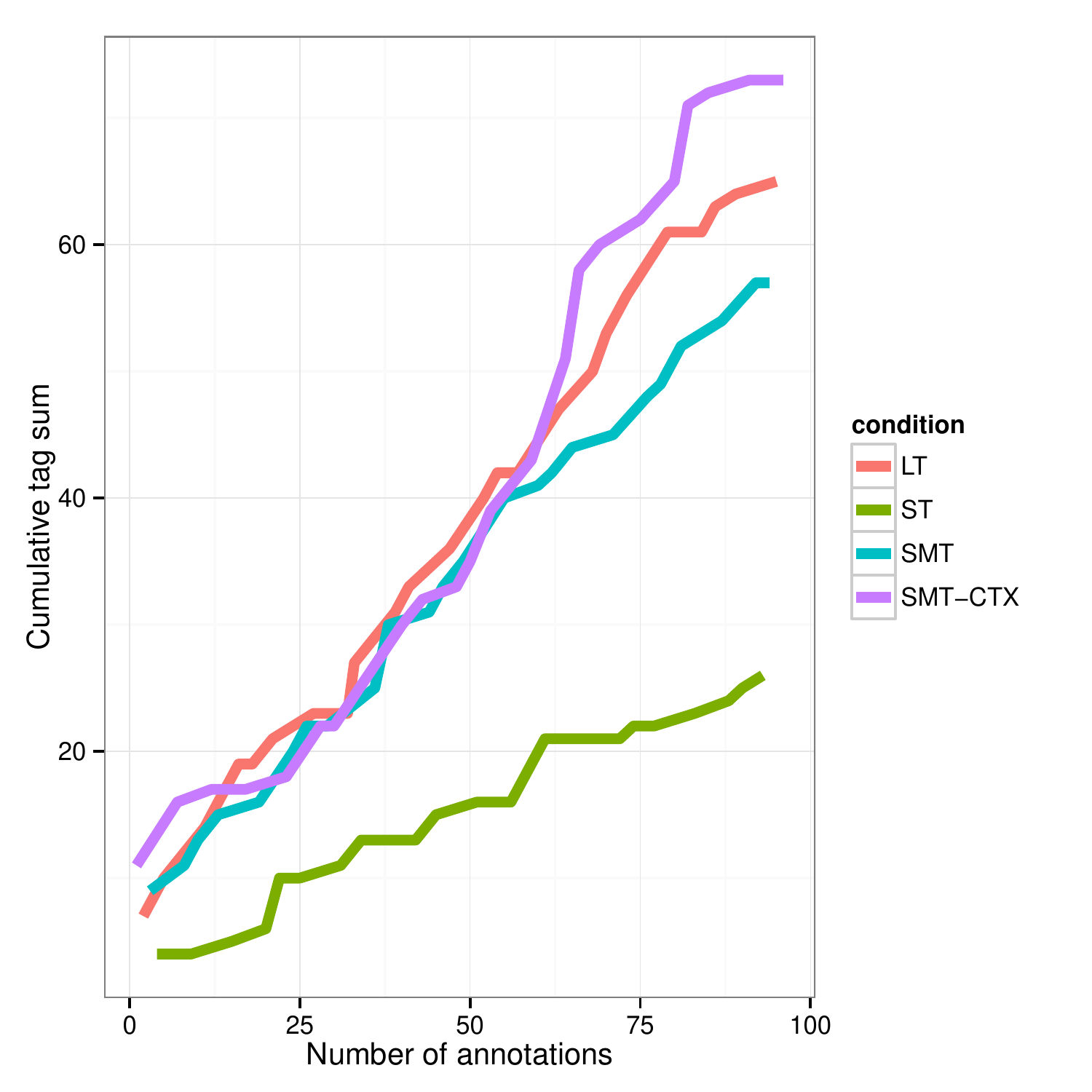}
    \caption{Tag evolution under different tagging conditions.}
    \label{fig_tag_evolution}
\end{figure}

Figure~\ref{fig_tag_evolution} shows the cumulative sum of tags grouped by condition and its evolution with the growing number of annotations in the system. It indicates that users confronted with condition ST added fewer tags at the beginning of the experiment and less than in other conditions throughout the experiment. The evolution of tags added through label-based tagging (LT) is roughly linear because this technique does not rely on any other contextual information. Even though semantic tagging (SMT, SMT-CTX) relies on the existence of concept definitions in external knowledge contexts (e.g., Wikipedia) it shows similar behavior than condition LT. The divergence from linearity starting approximately from annotation number 50 could be caused by the variance of the mean number of added tags as part of an annotation. However, more data is needed to verify this.

\subsection{Effect on Tag Types and Categories}

Our analysis of manually coded tags revealed that from all collected tags, 48\% were factual and 52\% personal for the context of a given map. Table~\ref{tab:tag_types} shows that the distributions of factual and personal tag assignments did not differ by condition, $\chi^2(3, N = 221) = 1.0516, p = .78$.

\begin{table}[ht]
  \centering
    {\renewcommand{\arraystretch}{1.2}%
    \begin{tabular}{rrrr}
        & Factual & Personal & \\ \otoprule%
      LT &  29 & 36 & 65\\ 
      ST &  14 & 12 & 26\\ 
      SMT &  29 &  28 & 57\\ 
      SMT-CTX &  33 & 40 & 73\\ 
      \bottomrule
        & 105 & 116 & 221 \\  
    \end{tabular}
    }
  \caption{Tag type contingency table.}
  \label{tab:tag_types}
\end{table}

The distribution of tag categories grouped by condition is shown in Table~\ref{tab:tag_categories} and reveals that locations (L) were tagged most frequently (54\%), followed by a large fraction (39\%) of tags classified as other (O). Only some tags (7\%)  referring to persons or groups (P), hardly any (2\%) event (E) and only one temporal tag has been added. From this we can conclude that users mainly contribute location-specific contextual information to maps when adding tags. As with tag types, the distributions of assigned tag categories did not differ by condition, $\chi^2(12, N = 221) = 17.30, p = .14$.

\begin{table}[ht]
  \centering
    {\renewcommand{\arraystretch}{1.2}%
    \begin{tabular}{rrrrrrr}
      & E & L & O & P & T \\  \otoprule%
      LT &   4 &  38 &  17 &   5 &   1 & 65\\ 
      ST &   0 &  14 &  11 &   1 &   0 & 26\\ 
      SMT &   1 &  33 &  17 &   6 &   0 & 57\\ 
      SMT-CTX &   0 &  34 &  35 &   4 &   0 & 73\\
      \bottomrule
        & 5 & 119 & 80 & 16 &  1 & 221\\
    \end{tabular}
    }
  \caption{Tag category contingency table.}
  \label{tab:tag_categories}
\end{table}

\subsection{Effect on User Task Load}

In the post-test survey we asked participants to (i) express their perceived task load in a TLX questionnaire and (ii) to rank the conditions they had seen according to intuitiveness, influence on annotation text, mental effort, and overall usefulness. Since we administered the TLX rating in an unpaired variant, each factor was rated individually and all factors were rated at once immediately after a participant was confronted with a condition.

\begin{figure}
  \centering
  \includegraphics[width=1\columnwidth]{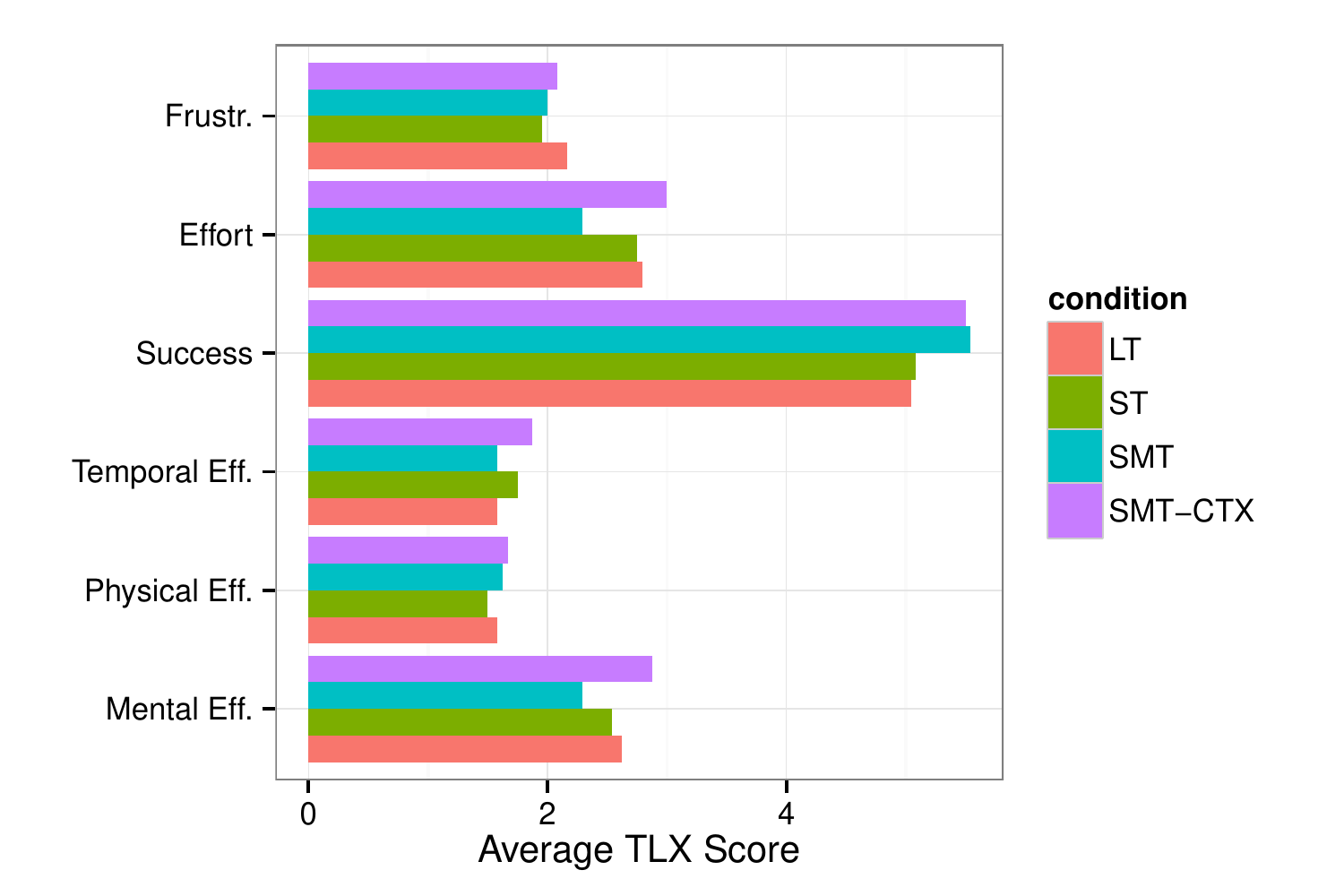}
  \caption{Average TLX scores per category and tagging condition.}
  \label{fig:tlx-summary}
\end{figure}

The results from the TLX questionnaire are summarized in Figure~\ref{fig:tlx-summary}. One-way analysis of variances conducted for each task load category indicated that there was no significant effect of the tagging conditions on frustration, F(3,92) = 0.095, p = .96, overall effort F(3,92) = 1.03, p = .38, success, F(3,92) = 0.687, p = .56, temporal effort, F(3,92) = 0.47, p = .70, physical effort, F(3,92) = 0.11, p = .95, and mental effort, F(3,92) = 0.75, p = .52. From this we can conclude that (i) the additional effort caused by semantic tagging techniques (SMT, SMT-CTX) has no effect on user task load, and (ii) there is no significant difference between semantic tagging with and without provided contextual information with respect to user task load.

The condition ranking scores results are summarized in Table~\ref{tab:preference}. They are based on the number of times a certain condition was voted first, second, etc. among the 24 participants. A Friedman rank sum test performed on the ordering task revealed significant differences for all conditions in all categories ($p < .01$).

% the output of this table was generated with the xtable R package
% e.g. xtable(t(table(data[,c("condition", "order_usefulness")])))
% "mental effort" is inversed
\begin{table*}
\begin{center}
\setlength{\tabcolsep}{4pt}
\resizebox{1.7\columnwidth}{!}{
\begin{tabular}{rrrrrrrrrrrrrrrrr}
  \toprule
  & \multicolumn{4}{c}{\textbf{Intuitiveness }} & \multicolumn{4}{c}{\textbf{Influence}} & \multicolumn{4}{c}{\textbf{Mental Eff.}} & \multicolumn{4}{c}{\textbf{Usefulness}}\\
  \cmidrule(r){2-5} \cmidrule(r){6-9} \cmidrule(r){10-13} \cmidrule(r){14-17}
       Rank   & LT  & ST  & SMT & CTX & LT & ST  & SMT & CTX  & LT  & ST  & SMT & CTX & LT   &  ST & SMT & CTX \\ 
  \cmidrule(r){1-1} \cmidrule(r){2-5} \cmidrule(r){6-9} \cmidrule(r){10-13} \cmidrule(r){14-17}
            1 &  1  &   6 &   5 &  \textbf{12}  &  2 &   6 &   4 &  \textbf{12}  &  \textbf{20} &   2 &   0 &   2 &    0 &   3 &   4 &  \textbf{17} \\  
            2 &  3  &   7 &   9 &   5  &  2 &   6 &  11 &   5  &   1 &   5 &   8 &  10 &    0 &  10 &   8 &   6 \\  
            3 &  7  &   8 &   4 &   5  &  2 &  10 &   8 &   4  &   0 &   3 &  15 &   6 &    2 &  11 &  10 &   1 \\  
            4 &  13 &   3 &   6 &   2  & 18 &   2 &   1 &   3  &   3 &  14 &   1 &   6 &   22 &   0 &   2 &   0 \\  
   \bottomrule
\end{tabular}
}
\end{center}
\caption{Rankings for all conditions. The values indicate how often a condition was ranked first, second, etc.}
\label{tab:preference}
\end{table*}   

\begin{itemize}
  \item \textbf{Intuitiveness:} SMT-CTX was voted the most intuitive condition, with SMT and ST scoring second and third place, respectively. On the other side of the scale, condition LT was judged to be the least intuitive.
  \item \textbf{Influence on Annotations:} Condition SMT-CTX was ranked first among the majority of our participants as being the most influential on the resulting annotation text, followed by SMT and ST. Condition LT was the least influential. This can be explained by the fact that our semantic tagging approach with additional context information might have offered inspiration to write more annotation text, as was confirmed by the comments participants gave us.
  \item \textbf{Mental effort:} Condition LT appeared to require the least effort, with conditions SMT-CTX, SMT and ST following. We assume that this result is based on the need to check suggested tags for their validity in the context of the annotation text that was already entered, whereas in condition LT, the users could only manually enter tags.
  \item \textbf{Usefulness:} Of all conditions, semantic tagging with context (SMT-CTX) was ranked as being the most useful, followed by ST and SMT. Condition LT was considered the least useful.
\end{itemize}

From this we can conclude that our participants found our semantic tagging user interface, showing contextual information, most useful and intuitive. One possible explanation of why the participants stated that semantic tagging influenced their annotation text is that Maphub suggested tags on-the-fly while users were entering their text.

\subsection{Other observations}

% annotation in general
The notes taken by the experiment administrators during each session reveal that several participants were not familiar with the concept of annotations, which confirms our pre-test survey findings. Looking at other peoples' annotations to get ideas was a commonly observed behavior among the participants. One participant explicitly stated that he was doing so because he wanted to annotate areas on the maps that were still untouched.

Many participants annotated places they are familiar with in real-life, such as cities they lived in. One participant, for example, explicitly said:

\begin{quotation}
  \emph{``I am doing this [annotation] because I am from the Philadelphia area''}
\end{quotation}

% tag suggestion
 
The participants generally liked the tag suggestion conditions, but expressed that their recommendation strategies should be improved: one participant meant that suggested and other users' tags were not right; others complained that too many non-relevant tags were suggested; one was confused because of non-relevant tag suggestions. She received a tag suggestion ``India'' for an annotation created on ``South Africa'' because another user added that tag as part of an annotation. Some participants rejected tags that experiment administrators thought were relevant. One participant wanted to add tags that better reflect personal concepts. Two participants didn't notice that there were different tagging conditions even though they were informed about this in the experiment instructions. One participant was ``'freaked out'' when tags referring to his personal status were suggested (e.g., Non-resident Indian).

% semantic tagging context

Participants reacted very differently to the semantic tagging with context (SMT-CTX) condition: Some did not notice the context information Maphub provided or did not hover over the tags very often; others explicitly asked for context when none was provided or hovered over each suggested tag long enough to see and read the Wikipedia excerpt. One participant found this feature very useful and wanted to click the appearing context notes to see the entire Wikipedia page.

% effect on annotations

Another commonly observed behavior was that people who made spelling errors (e.g., on city names) purposely altered their annotation text to retrieve corresponding semantic tags. Tag suggestions also motivated them to type more text, and the administrators observed that participants interrupted writing to see tags appearing.

% usability issues

Several participants suggested general usability improvements for Maphub. Being able to move, close and reopen an annotation dialog window to look at the map underneath was thought to be useful. Some also had troubles drawing annotation shapes on the trackpad of the laptop we used in our experiment. Some participants thought that a combination of label-based and semantic tagging would be the most useful tagging system combination.

% general

Even though few of the participants had expressed particular interest in history and historic maps, most of them became interested in those subjects during the experiment session and spent considerable time viewing the maps. One participant said:

\begin{quotation}
  \emph{``I love maps so much; the old maps are so interesting because it shows you the way people thought about the world.''
}\end{quotation}

Many participants were excited when they found places they knew. They were entertained when they noticed map features they wouldn't be able to find on modern maps, such as drawings of elephants and dragons on the Waldseem\"{u}ller map.

\section{Discussion}\label{sec:discussion}

In our study we compared semantic tagging without (SMT) and with (SMT-CTX) context against label-based tagging (LT) and suggestive tagging based on tags added by other users (ST).

\subsection{Findings}

% Tag usage results
Our empirical evaluation revealed that semantic tagging leads to at least to same number of positive (accepted) tagging relationships as label-based tagging, with the main advantage being that semantic tagging relationships are qualified and unambiguously refer to user-chosen and well-defined concepts such as the Wikipedia article about ``Paris''. We believe that semantic tagging is more accurate than any a-posteriori mapping technique that tries to associate tag labels to concept definitions, since only the user who creates a tag will know its precise semantics.

When comparing the suggestive techniques only, both semantic tagging techniques (SMT, SMT-CTX) lead to more positive (accept) and less negative (reject) tagging relationships than suggestive tagging based on other users' tags. This demonstrates that taking a broader knowledge context such as Wikipedia as a basis for tag recommendation is a reasonable, complementary approach to drawing tags from other users, who might have added a large fraction of personal tags that do not apply for other users.

A major advantage of any suggestive tagging technique with user feedback is that negative (reject) tagging relationships can also be recorded. It allows, for instance, a user to say that a certain image is about ``Paris (the city)'' but not ``Paris (the character in Greek mythology)''. Machine learning techniques could exploit this knowledge and improve tag recommendation, which was also suggested by many participants in our experiments. We believe that semantic tagging might lead to more balanced tagging graphs than other suggestive techniques.

Semantic tagging is also a possible solution to overcome the tag recommendation bootstrap problem, which occurs when a system has not yet collected sufficient tags for providing meaningful tag recommendations. It might also help users who do not add tags because they simply could not think of any, and generally encourage more people to use tags, as pointed out by Sen et al.~\cite{sen2006tagging}.

% Tag type and category results
The fraction of personal, non-factual tags we collected in our experiments is higher than in other studies (e.g.,~\cite{sen2006tagging,Sigurbjornsson:2008vn}). We believe that is because of the historical map subject of our study, which motivated many participants to add annotations related to their personal context, such as places where they lived or attend(ed) school. As a result, annotations contained the school's name (e.g., Cornell University) or activities conducted there (e.g., played Tennis) and corresponding tags were added. However, our results also show that this is not affected by the tagging condition.

The categories of tags added in our experiment were mostly related to some location. This follows the findings of Sigurbj\"ornsson et al.~\cite{Sigurbjornsson:2008vn} and is comprehensible in an experiment focusing on the annotation of geography-related material, such as historic maps.

% User Task Load
The results of our user task load survey also show that the semantic tagging feature, with and without context, can be implemented by systems without increasing the users' task load compared to label-based tagging interfaces.

\subsection{Limitations}

The small sample sizes (maps, participants) are clear limitations of our study and prevent us from examining and comparing the distribution of tag frequencies with findings from other studies, which showed that collaborative tagging systems tend to follow a power-law distribution~\cite{Halpin:2007kx}. We also used only a single, very broad knowledge context (Wikipedia) in our experiment and can therefore not generalize these results for other application scenarios and other contexts. Applying semantic tagging in a more domain-specific setting with a narrower knowledge context might lead to less proposed and therefore less accepted tags.
 
Previous studies (e.g.,~\cite{ames2007we,gupta2010survey}) identified two major tagging motivations: (i) easing later personal retrieval and (ii) making content findable by others. Our results confirm this also for semantic tagging. However, the efficiency of semantic tagging for personal retrieval depends largely on the applied knowledge context. Specific vocabularies, such as a gazetteer for historical place names, might not define the concepts user would apply for personal knowledge organization. But it is certainly possible to combine semantic tagging with label-based tagging to serve both types of user motivation.

Performance of existing named entity recognition services was a technical limitation. These services also support named entity disambiguation, but only at the cost of higher computational complexity and reduced response time. This is undesirable in the context of semantic tagging, which requires fast response times and allows lower precision because it lets the user decide on the meaning of an entity. We were able to overcome this limitation by deploying Wikipedia miner on a high memory machine (20 GB RAM) and reducing disambiguation thresholds to zero.

\subsection{Practical Implications}

Overall, we believe that our findings carry implications for designers who want to adopt semantic tagging in other scenarios. A major incentive for system providers to implement tagging is to obtain metadata describing the content and context of online resources, which is important for efficient resource discovery but expensive in terms of time and effort when created manually~\cite{Duval:2002fk}. In traditional, label-based tagging systems providers can add possibly ambiguous label-based tags to their records. With semantic tagging, they obtain references to concepts defined in other Web-based knowledge context. Traditional information retrieval techniques can be enhanced to exploit these relationships and consider additional contextual information.

\section{Conclusion}\label{sec:conclusion}

% Problem
In this paper we discussed the problem of semantically ambiguous tags, which are generated by current tagging systems. We proposed semantic tagging as a possible solution in which a user adds unambiguous URI references drawn from some knowledge context to shared content. We applied this general technique in the specific context of historical maps and allow users to annotate and tag them.
% Important Results

Our semantic tagging implementation does not affect tag production, the types and categories of obtained tags, and user task load while providing tagging relationships to well-defined concept definitions. When compared to label-based tagging, our technique also gathers positive and negative tagging relationships. It does not affect tag type and category distributions and can lead to higher user satisfaction. 

% Future work
We aim at extending our study along several dimensions: first we would like to run another experiment with expert users (e.g., library professionals) and compare these findings with our current ones. Second, we would like to deploy the semantic tagging feature in a larger system to generalize our findings. Third, we would like to work on tag suggestion techniques that better exploit the availability of positive and negative tags. And fourth, we would like to continue developing Maphub and encourage more users to share their stories and knowledge they might have about historical maps.

\section{Acknowledgments}

This work is supported through a generous grant from the Andrew W. Mellon Foundation and by a Marie Curie International Outgoing Fellowship within the 7th European Community Framework Programme (PIOF-GA-2009-252206).

% Balancing columns in a ref list is a bit of a pain because you
% either use a hack like flushend or balance, or manually insert
% a column break.  http://www.tex.ac.uk/cgi-bin/texfaq2html?label=balance
% multicols doesn't work because we're already in two-column mode,
% and flushend isn't awesome, so I choose balance.  See this
% for more info: http://cs.brown.edu/system/software/latex/doc/balance.pdf
%
% Note that in a perfect world balance wants to be in the first
% column of the last page.
%
% If balance doesn't work for you, you can remove that and
% hard-code a column break into the bbl file right before you
% submit:
%
% http://stackoverflow.com/questions/2149854/how-to-manually-equalize-columns-
% in-an-ieee-paper-if-using-bibtex
%
% Or, just remove \balance and give up on balancing the last page.
%
\balance

% If you want to use smaller typesetting for the reference list,
% uncomment the following line:
% \small
\bibliographystyle{acm-sigchi}
\bibliography{references}
\end{document}